\documentclass[sigconf, 10pt]{acmart}

\usepackage[english]{babel}
\usepackage{blindtext}

\usepackage{xspace}
\usepackage{enumitem}
\usepackage{graphicx}
\usepackage{gensymb}
\usepackage{booktabs}
\usepackage{hyperref}
\usepackage{subcaption}
\usepackage{multirow}
\usepackage{multicol}
\usepackage{bbding}
\usepackage{adjustbox}
\usepackage{hyperref}
\usepackage[noend]{algpseudocode}
\usepackage{algorithmicx}
\usepackage{algorithm}
\usepackage{amsmath}
\usepackage{multirow}  
\usepackage{enumerate}
\usepackage{tabularx}
\usepackage{xcolor}
\definecolor{custompink}{HTML}{FD7272}
\definecolor{customorange}{HTML}{F28E2B}
\definecolor{coolblue}{HTML}{2D98DA}
\definecolor{freshgreen}{HTML}{20BF6B}

\newcommand{\sysname}{\textit{ECCO}\xspace}
\newcommand{\myparagraph}[1]{\vspace{1mm} \noindent \textbf{\textsf{#1}}}

\newcommand{\blue}[1]{\textcolor{black}{#1}}

\renewcommand\footnotetextcopyrightpermission[1]{} 
\setcopyright{none}

\acmDOI{}

\acmISBN{}

\acmPrice{}

\begin{document}
\title{ECCO: Leveraging Cross-Camera Correlations for Efficient Live Video Continuous Learning}
\author{Yuze He$^{\dagger}$, Ferdi Kossmann$^{*}$, Srinivasan Seshan$^{\dagger}$, Peter Steenkiste$^{\dagger}$}

\affiliation{$^{\dagger}$Carnegie Mellon University \hspace{3mm}
$^{*}$Massachusetts Institute of Technology }

\renewcommand{\shortauthors}{X.et al.}

\begin{abstract}
Recent advances in video analytics address real-time data drift by continuously retraining specialized, lightweight DNN models for individual cameras. However, the current practice of retraining a separate model for each camera suffers from high compute and communication costs, making it unscalable. We present \sysname, a new video analytics framework designed for \textit{resource-efficient} continuous learning. The key insight is that the data drift, which necessitates model retraining, often shows temporal and spatial correlations across nearby cameras. By identifying cameras that experience similar drift and retraining a shared model for them, \sysname can substantially reduce the associated compute and communication costs. Specifically, \sysname introduces: (i) a lightweight grouping algorithm that dynamically forms and updates camera groups; (ii) a GPU allocator that dynamically assigns GPU resources across different groups to improve retraining accuracy and ensure fairness; and 
(iii) a transmission controller at each camera that configures frame sampling and coordinates bandwidth sharing with other cameras based on its assigned GPU resources. 
We conducted extensive evaluations on three distinctive datasets for two vision tasks. 
Compared to leading baselines, \sysname improves retraining accuracy by $6.7\%$-$18.1\%$ using the same compute and communication resources, or supports $3.3\times$ more concurrent cameras at the same accuracy.
\end{abstract}

\maketitle
\pagestyle{plain} 
\vspace{-0.5em}
\section{Introduction}
\label{sec:intro}
The rapid expansion of camera deployments \cite{camera1, camera2, camera3} is driving a growing demand for live video analytics, with the market expected to reach 22.6 billion USD by 2028 \cite{market}. Live video analytics uses deep neural network (DNN) models to perform vision tasks such as object detection and classification. These analytics are at the core of applications in diverse fields like enterprise security \cite{security}, traffic monitoring \cite{traffic}, and autonomous driving \cite{ad}. To process live video streams in real time and ensure low-latency results, it is often crucial to deploy DNNs and run inference directly on edge devices \cite{edge1, edge2}. 
However, since edge devices often have limited resources (with less powerful GPUs \cite{aws, azure}), these devices typically use lightweight, specialized models instead of complex, generic models \cite{compress1, compress2}.

\begin{figure}
    \centering
    \includegraphics[width=\linewidth]{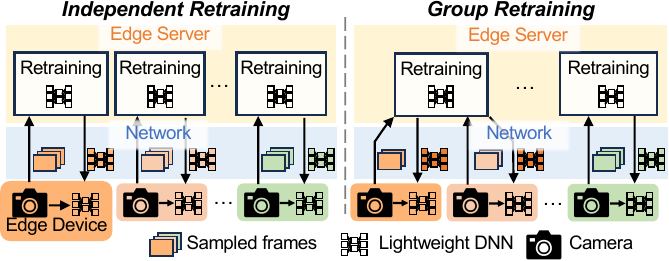}
    \vspace{-2em}
    \caption{Frameworks of existing continuous retraining systems (left) and our proposed system (right).}
    \vspace{-1.5em}
    \label{fig:teaser}
\end{figure}
These lightweight models are initially trained using representative data from each camera. Once deployed, these models face challenges from \textit{data drift}, where the live video data changes significantly from the initial training data. For example, cameras deployed in urban areas or mounted on moving vehicles may record changes in object types, activities, lighting conditions, or crowd densities over time. As a result, the accuracy of the DNN's predictions can substantially decline. A promising solution to tackle data drift is \textit{continuous learning}. This approach retrains the lightweight models on an edge server using more recent video frames transmitted from the cameras, helping the models to adapt to new data patterns. Recent studies \cite{lvacv1, lvacv2} have shown the benefits of continuous learning in enhancing the robustness and accuracy of video analytics systems.

A key challenge in continuous learning is resource efficiency, as practical deployments are often constrained by compute and communication resources. Two critical resources in this setting are the GPUs at the edge server and the bandwidth between the distributed cameras and the server. Prior works have only focused on improving GPU efficiency (i.e., model accuracy per GPU unit). For example, Ekya \cite{ekya} and AdaInf \cite{adainf} optimize GPU scheduling across retraining (and inference) tasks from multiple cameras. RECL \cite{recl} enhances GPU efficiency by reusing historical models as starting points for retraining. However, these methods have two significant limitations. First, they largely overlook bandwidth efficiency, which is related to GPU usage and should be optimized jointly. 
Second, all these systems assume training a separate model for each camera, a strategy we refer to as ``\textit{independent retraining}'' (see Fig.~\ref{fig:teaser} (left)). This design can result in redundant GPU computation, especially when cameras exhibit correlated data patterns. 

\myparagraph{Core idea:} We introduce \textit{group retraining}, a new approach that groups cameras experiencing similar data drift and retrains a shared model using their collective data (see Fig.~\ref{fig:teaser} (right)). The rationale for group retraining is two-fold. First, data drift often exhibits temporal and spatial correlation among some cameras, e.g., traffic cameras at the same intersection or cameras mounted on vehicles traveling together may encounter similar environmental changes. Second, lightweight models, despite their compact architecture, can generalize across similar data distributions and may even benefit from subtle variations observed by different cameras \cite{survey1, survey2}. By retraining one model for a group of cameras instead of individual models for each, we reduce the number of models that need retraining. This reduces compute costs and lowers the data transmission requirements for each camera to the edge server.

\myparagraph{Technical challenges:}
Realizing the potential of group retraining involves addressing several system-level challenges. 
First, efficiently identify cameras with similar data drift is challenging because video streams are high-dimensional, making direct cross-camera comparisons computationally expensive. Moreover, as live video content changes continuously, camera grouping cannot be a one-time operation and must be updated over time.
Second, allocating GPU resources efficiently across camera groups is non-trivial. As we will show in \S\ref{sec:schedule}, naive extensions of existing GPU allocation algorithms are ill-suited for group retraining. They tend to favor larger groups while under-provisioning smaller ones, resulting in unfairness among cameras.
Third, in addition to GPU resources, network bandwidth is a critical yet often overlooked constraint. Since different groups’ models are retrained using live data transmitted from their distributed cameras, bandwidth allocation across groups must be coordinated with GPU allocation to maximize retraining efficiency. This coordination is challenging because it often requires assigning unequal bandwidth shares to different groups, which may conflict with the equal-share behavior enforced by standard congestion control mechanisms.

\myparagraph{Our solution:} We introduce \sysname, a new continuous retraining framework for live video analytics that leverages cross-camera correlations to improve resource efficiency, scalability, and responsiveness. \sysname addresses the above challenges through three key strategies:
(i) A lightweight grouping algorithm that first narrows down candidate cameras using metadata (e.g., drift time and location), and then makes grouping decisions by evaluating the accuracy gain.
(ii) A new formulation of the GPU allocation problem tailored for group retraining, along with a GPU allocator that jointly optimizes for overall performance and fairness across groups.
(iii) A transmission controller at each camera that adapts frame sampling (data volume) to match its assigned GPU resource and adjusts its transmission rate using a customized congestion control algorithm, that enables bandwidth allocation across groups in proportion to their GPU shares in a best-effort manner.

We implemented and evaluated \sysname on two computer vision tasks: object detection and instance segmentation, and compared it with state-of-the-art video analytics systems for three datasets. Using the same compute and communication resources, \sysname improves the mean Average Precision (mAP) by $6.7\%$-$16.6\%$ for object detection and $9.3\%$-$18.1\%$ for instance segmentation over strong baselines. \blue{While this improvement may appear to be a small change in accuracy, in practice this improvement translates to supporting $3.3\times$ more cameras at the same accuracy level.} We also show that \sysname’s GPU allocator optimizes both performance and fairness, which significantly reduces the accuracy gap across groups while maintains comparable overall accuracy to the baseline allocator. The transmission controller improves bandwidth efficiency, requiring only $25\%$-$33\%$ of the bandwidth used by baselines to reach similar accuracy with the same GPU budget. Another finding is that \sysname's advantage in responsiveness becomes more pronounced under low-bandwidth conditions, reducing retraining latency by more than $5\times$ due to group retraining’s effective data aggregation and natural model reuse within the group.

\begin{figure*}[]
	\begin{minipage}[b][][b]{\textwidth}
		\centering
		\includegraphics[width=1\textwidth]{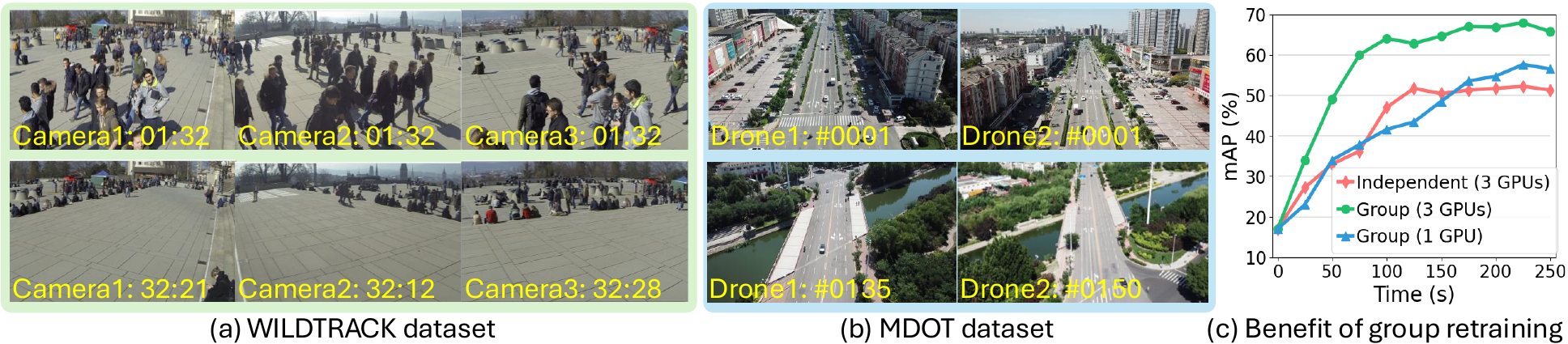}\\[-1ex]
	\end{minipage}
    \vspace{-2em}
	\caption{A motivation study. (a) \& (b): Example frames from two public datasets showing that data drift exhibits significant temporal and spatial correlations among cameras. (c) Model accuracy of three retraining settings.}
\vspace{-1em}
	\label{fig:moti}
\end{figure*}
\section{Motivation and Observations}
\label{sec:motivation}
\subsection{Limitations of Independent Retraining}
\label{sec:cl_intro}
To highlight the inefficiencies of current retraining approaches, we first revisit the standard retraining pipeline shown in Fig.~\ref{fig:teaser} (left). Edge devices perform real-time inference on live video using local lightweight models (``students''). When data drift is detected\footnote{Several prior works~\cite{foot1,foot2,foot3} have been proposed to detect data drift or scene changes in video streams and can serve as retraining triggers.}, the edge device sends a retraining request to the edge server and begins continuously sampling and transmitting video frames as retraining data. The server uses a high-accuracy but resource-intensive model (``teacher'') to annotate these frames, retrains a separate student model for each camera using its own data, and returns the updated model to the corresponding device.

This design faces an obvious scalability challenge: as the number of cameras increases, the retraining workload grows linearly, placing significant strain on both GPU and bandwidth resources, which increases retraining latency. Moreover, when multiple cameras experience similar data drift, independently retraining separate models becomes redundant and inefficient.

\vspace{-0.5em}
\subsection{Why Group Retraining?}
\label{sec:case_study}
\myparagraph{Cross-camera correlations in data drift.}
The main insight motivating our work is that data drift among cameras can exhibit spatial and temporal correlations. Cameras that are geographically close often experience similar environmental changes at similar times. Fig.~\ref{fig:moti}(a) and (b) illustrate examples of such drift using two representative camera types: static cameras from the WILDTRACK~\cite{wildtrack} dataset and mobile cameras from the MDOT~\cite{mdot} dataset. In WILDTRACK, static cameras monitoring a plaza show similar changes in foreground content (e.g., pedestrian density and behavior). In MDOT, mobile cameras mounted on drones flying in formation capture similar shifts in both foreground (e.g., vehicle densities) and background (e.g., buildings and vegetation) as they move through urban and suburban areas. This phenomenon is also observed in other public datasets such as Cityflow~\cite{cityflow}, Bellevue Traffic Video~\cite{bellevue}, VERI-Wild~\cite{veriwild}, and DukeMTMC~\cite{dukemtmc}. These examples suggest that independent retraining can be inefficient when data distributions across cameras are closely aligned.

We propose a new approach, \textit{group retraining}, which aggregates retraining requests from cameras with correlated data drift and uses their collective data to retrain a single shared model. While prior work has used inter-camera correlations for tasks such as object re-identification~\cite{spatula} or video inference configuration~\cite{chameleon}, our approach is, to our knowledge, the first to exploit these correlations to reduce the cost of model retraining.

\myparagraph{Benefits of group retraining.} 
To demonstrate the potential benefits of group retraining, we conduct a case study using three drone videos selected from the MDOT dataset, identified as correlated based on manual inspection. These videos exhibit similar scene changes as the drones fly in formation, resembling the scenario illustrated in Fig.~\ref{fig:moti}(b). We use a high-performance object detection model, YOLO11x (194.9 BFLOPs), as the teacher and a lightweight version, YOLO11n (6.5 BFLOPs), as the student model. We compare three retraining settings: (i) Independent retraining: Each camera retrains its own model using 1 GPU (3 GPUs in total). (ii) Group retraining (3 GPU): A shared model is trained using data from all three cameras with 3 GPUs. (iii) Group retraining (1 GPU): The shared model is retrained using the same data with only 1 GPU.

Fig.~\ref{fig:moti}(c) reports object detection accuracy over time, measured by mean Average Precision (mAP) averaged across the three cameras. We observe that: (i) With the same GPU resources, group retraining (green) achieves higher accuracy and lower retraining latency than independent retraining (red). (ii) Even with only 1 GPU, group retraining performs comparably to independent retraining using 3 GPUs. These results suggest that group retraining can significantly reduce retraining cost in GPU use, while improving responsiveness.
\section{Design of \sysname}
\label{sec:design}
This paper proposes \sysname, a continuous learning system for live video analytics that improves compute and communication efficiency by leveraging cross-camera correlations through group retraining.

\begin{figure}[]
    \centering
    \includegraphics[width=\linewidth]{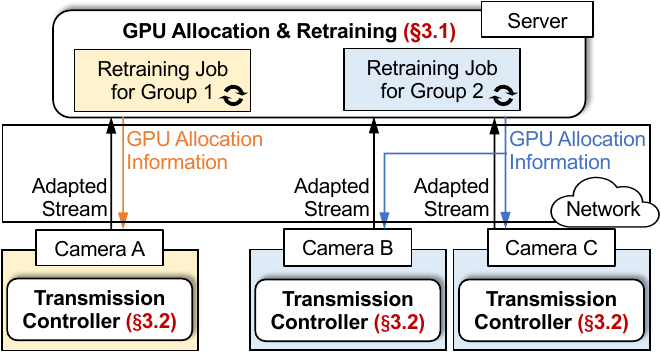}
    \vspace{-2em}
    \caption{Steady state of \sysname. The server-side GPU allocator and camera-side transmission controllers coordinate to support group retraining. Colors indicate different groups.}
    \vspace{-1.5em}
    \label{fig:steady}
\end{figure}

\begin{figure}[t]
    \centering
    \includegraphics[width=\linewidth]{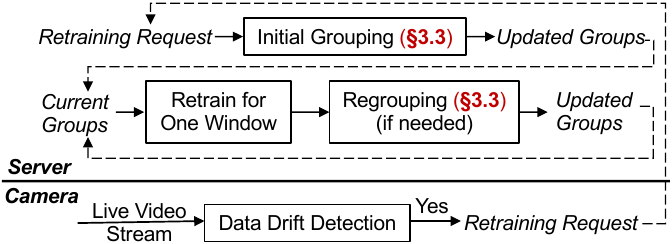}
    \vspace{-2em}
    \caption{Dynamic camera grouping in \sysname. Initial grouping is triggered by retraining requests, while periodic regrouping updates groups during retraining.}
    \vspace{-1.5em}
    \label{fig:update}
\end{figure}

\myparagraph{Overall architecture (Fig.\ref{fig:steady} and Fig.\ref{fig:update}):}
\sysname comprises a server, multiple edge devices (cameras), and the network connecting them. We make no assumptions about the underlying network connectivity.


Fig.\ref{fig:steady} illustrates steady-state operations, where cameras have already been grouped based on data drift similarity. Each group’s retraining is handled by a single job. Retraining is managed in discrete retraining windows, which serve as the basic unit for coordination and resource management. Within each retraining window, \sysname improves resource efficiency and retraining accuracy through two coordinated modules.

The GPU allocator at the server dynamically distributes GPU resources across groups, guided by an objective that balances overall retraining performance with fairness among groups (\S\ref{sec:schedule}). The GPU allocation information is transmitted to cameras to guide their transmission strategies.

To maximize retraining accuracy, the transmission controller at each camera configures training data and regulates transmission to match the allocated GPU budget. First, it selects a sampling configuration (frame rate and resolution) based on both the GPU budget and the observed scene characteristics (\S\ref{sec:sampling}). Second, it aligns its bandwidth usage with the assigned GPU share. This is achieved using a customized GAIMD congestion control algorithm \cite{gaimd}, which enforces flows to compete with controlled aggressiveness proportional to their GPU share, thereby approximating GPU-proportional bandwidth allocation under network constraints (\S\ref{sec:bwallocation}). While non-AIMD approaches are possible, they are not the focus of our efforts.

Fig.\ref{fig:update} illustrates how camera groups are created and updated in response to data drift. Camera grouping involves two stages. The first is initial grouping, triggered when a camera detects drift and initiates a retraining request. At this point, the server determines whether the camera should join an existing group that targets similar data drift or form a new group. The second is regrouping, performed periodically on existing groups during retraining. This process adapts group membership to evolving data distributions in live video streams, as some cameras may gradually drift and no longer remain similar to others in their current group.

\vspace{-0.5em}
\subsection{GPU Allocation for Group Retraining}
\label{sec:schedule}
To scale and serve more cameras, \sysname must efficiently allocate GPU resources across camera groups. Unlike existing allocation methods designed for independent retraining \cite{adainf,recl,ekya}, group retraining introduces new challenges that those methods do not address. We thus formulate a new optimization objective that balances overall retraining accuracy with fairness across groups. To achieve this, we design a GPU allocation algorithm that tracks each group’s current accuracy and its accuracy improvement with added compute resources. It then dynamically allocates more GPU resources to groups that either have low accuracy or show larger improvement when given more GPU resources.

\myparagraph{Limitations of existing approaches:} Prior work typically allocates GPU resources to maximize total accuracy improvement (or average accuracy) across all cameras over a retraining window. However, directly applying this strategy to group retraining introduces bias: it implicitly favors larger groups, since improvements from a group with more cameras contribute more to the global accuracy sum.

To illustrate this issue, consider two groups: G1 with four cameras and G2 with one. If given equal GPU time, G1's model accuracy improves by $10\%$, and G2's by $15\%$. Since G1’s improvement benefits four cameras, its total contribution is $4 \times 10\% = 40\%$, while G2’s is only $15\%$. Thus, existing algorithms \cite{ekya,recl} tend to allocate most GPU resources to G1, leading to prolonged starvation for G2---until the training efficiency of G1 (i.e., accuracy gain per GPU second) naturally declines as training converges. The root cause is that existing objectives implicitly weight groups by size, which is suitable for independent retraining \blue{but introduces a systematic allocation bias against smaller groups in the group retraining setting.} 

\myparagraph{Objective reformulation:} To address this, we define a new objective that balances system-wide accuracy and fairness across groups. Consider a set of retraining jobs $\mathcal{J}$, one per group, running on $G$ GPUs over a retraining window $T$ of duration $\|T\|$, with total compute capacity $G\|T\|$ GPU-time. Let $A_j(g_j)$ be the accuracy of group $j$'s model after receiving $g_j$ GPU-time, averaged over its $n_j$ camera members. We optimize:
\begin{equation}
\begin{aligned}
\max_{\{g_j\}_{j \in \mathcal{J}}}  \left[ \alpha\frac{\sum_{j \in \mathcal{J}} n_j^{\beta} A_j(g_j)}{\sum_{j \in \mathcal{J}} n_j^{\beta}} + \min_{j \in \mathcal{J}} A_j(g_j) \right]
\text{ s.t.} \sum_{j \in \mathcal{J}} g_j \leq G\|T\|
\end{aligned}
\label{eq:objective}
\end{equation}
The first term captures the overall average accuracy across all groups, weighted by $n_j^\beta$, where $\beta \leq 1$ controls the influence of group size, i.e., how much distinction is made between groups of different sizes. The second term promotes fairness by maximizing the worst-performing group's accuracy. The parameter $\alpha$ adjusts the balance between the two goals. 
\begin{algorithm}[t]
\caption{GPU Allocation for Group Retraining} \label{alg:schedule}
\begin{algorithmic}[1] 
\State \textbf{Input:} retraining jobs \( \mathcal{J} \), retraining window size of \( W \) micro-windows, constants \( \alpha, \beta \)
\State \( \text{budget} \gets W \), initialize \( Acc[], AccGain[], ObjGain[] \) 
\Procedure{MicroRetraining}{j}
    \State \( \text{acc}_i \gets j.\text{EVAL}() \)
    \State Train job \( j \) for one micro-window
    \State \( \text{acc}_f \gets j.\text{EVAL}() \)
    \State \( \text{budget} \gets \text{budget} - 1 \)
    \State \( Acc[j] \gets acc_f \); \( AccGain[j] \gets acc_f - acc_i \)
\EndProcedure
\Procedure{CalObjectiveGain}{}
    \For{\( j \) in \( \mathcal{J} \)}
        \State \( ObjGain[j] \gets \frac{\alpha n_j^{\beta}}{\sum_{j \in \mathcal{J}} n_j^{\beta}} AccGain[j] \)
    \EndFor
    \State \( {ObjGain[\text{argmin}(Acc)]+=AccGain[\text{argmin}(Acc)]} \)
\EndProcedure
\For{\( j \) in \( \mathcal{J} \)}\Comment{\textbf{Initial training pass}}
    \State MicroRetraining($j$); \( CalObjectiveGain() \)
\EndFor
\State \text{Estimate per-group GPU resource (\S\ref{sec:rate_control})}
\While{\( \text{budget} > 0 \)}
    \State \( j \gets \text{argmax}(ObjGain) \)
    \State MicroRetraining($j$); \( CalObjectiveGain() \)
\EndWhile
\end{algorithmic}
\end{algorithm}

\myparagraph{GPU allocation algorithm:} We build on the resource allocation algorithm in \cite{recl} and propose a modified version tailored to our optimization objective. We time-share the GPU across multiple retraining jobs by dividing a retraining window $T$ into $W$ micro-windows. During each micro-window, one retraining job exclusively uses all GPUs. \blue{The key idea is to greedily assign GPU time to the group that yields the highest marginal improvement in the objective Eq.~\ref{eq:objective}.}

Algorithm~\ref{alg:schedule} outlines the procedure. It starts with an initial training pass (Lines 13–14), where each job trains for one micro-window to establish its short-term accuracy trajectory. After this, we measure each job's ``accuracy gain'', defined as the improvement in accuracy before and after training (Lines 3–8). This is then converted into a job-specific ``objective gain'' (Lines 9–12), which estimates its contribution to the overall objective (Eq.\ref{eq:objective}). For most jobs, the objective gain corresponds to the first term of Eq.\ref{eq:objective}, $\frac{\alpha n_j^{\beta}}{\sum_{j \in \mathcal{J}} n_j^{\beta}}AccGain$, capturing their marginal contribution to the weighted average accuracy. For the lowest-accuracy job, we also include the second term of Eq.~\ref{eq:objective}, giving $(\frac{\alpha n_j^{\beta}}{\sum_{j \in \mathcal{J}} n_j^{\beta}}+1)AccGain$, where the additional $AccGain$ serves as a fairness bonus to promote balance and prevent starvation. After this initialization, the algorithm repeatedly selects the job with the highest objective gain and assigns it the next micro-window (Lines 16–18). Objective gains are updated after each micro-window to reflect the latest performance trends. This greedy allocation continues until the GPU time budget is exhausted, approximating maximization of Eq.~\ref{eq:objective} under the GPU constraint.

\blue{\myparagraph{GPU allocation estimation for transmission control:} While the actual GPU allocation is performed dynamically as detailed above, the transmission controller at each camera (\S\ref{sec:rate_control}) requires an upfront estimate of its group’s expected compute budget to guide transmission. To provide this signal, the server uses the objective gains from the initial training pass to estimate each group’s GPU share for the current window (Line 15). Specifically, the estimated GPU resource $c_j$ for group $j$ is proportional to its objective gain relative to all groups: $ c_j = \frac{\text{ObjGain}[j]}{\sum_{i \in \mathcal{J}} \text{ObjGain}[i]} G \|T\|$. We also define $p_j=\frac{\text{ObjGain}[j]}{\sum_{i \in \mathcal{J}} \text{ObjGain}[i]}$ as the normalized GPU share weight for group $j$. The pair $(c_j, p_j)$ is then communicated to the group’s cameras as GPU allocation information.}

\vspace{-0.5em}
\subsection{Resource-Aware Transmission Control}
\label{sec:rate_control}
At the camera side, video transmission is governed by two external resources: GPU allocation and bandwidth availability. GPU allocation is decided by the server, which determines both (i) the group’s capacity to consume data and (ii) its relative priority for receiving more data. In practice, this capacity can be expressed as the maximum number of pixels per second that the GPU can process. Bandwidth availability, in contrast, is not explicitly allocated; it is constrained by network conditions and realized through congestion control.

Given these constraints, the camera controls three parameters of transmission: frame rate, resolution, and compression level. We refer to frame rate and resolution together as the \textit{sampling configuration}. \blue{Whenever the server provides updated GPU allocation information, the camera selects a sampling configuration whose frame rate–resolution product (pixels per second) stays within the GPU budget; multiple valid choices exist, and we discuss selection strategies in \S\ref{sec:sampling}.} During streaming, the camera then adjusts the compression level continuously to ensure the selected configuration can be delivered within the bandwidth actually achieved on the network. To approximate GPU-proportional weighted bandwidth allocation while respecting network constraints, we employ a customized GAIMD congestion control algorithm.
\subsubsection{Adaptation of Sampling Configuration}
\label{sec:sampling}
\blue{The sampling configuration impacts retraining quality because a limited GPU budget caps training throughput \blue{(e.g., total number of pixels processed per second \cite{pixel})}. As a result, increasing frame rate often requires lowering resolution, and vice versa. Retraining performance is therefore tightly coupled with how this tradeoff is managed under GPU constraints.}

\begin{figure}[t]
    \centering
    \includegraphics[width=\linewidth]{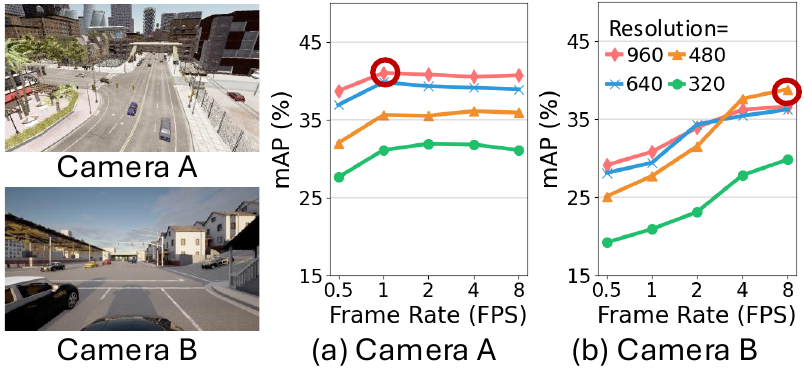}
    \vspace{-2em}
    \caption{\blue{Impact of sampling configurations (frame rate and resolution) on retraining accuracy under a fixed GPU budget for two distinct camera types. }}
    \vspace{-1.5em}
    \label{fig:moti_conf}
\end{figure}

To study this tradeoff, we conduct a case study using two representative camera types: a static, high-mounted traffic camera (A) and a mobile, vehicle-mounted camera (B), both simulated in the CARLA~\cite{carla} autonomous driving simulator (see Fig.~\ref{fig:moti_conf}). Using the model setup in \S\ref{sec:case_study}, we retrain a model for each camera using various sampling configurations while keeping the GPU budget fixed. To ensure a fair comparison, we fix the transmission bitrate to $1\,Mbps$ across all settings. This isolates the impact of sampling choices under consistent GPU and bandwidth conditions. Fig.~\ref{fig:moti_conf} shows the resulting retraining accuracy across configurations.

Two observations emerge. First, with the same GPU budget, retraining accuracy varies significantly across sampling configurations—by up to $2\times$. Second, the optimal configuration (circled in the figure) varies across camera types. This variation stems from intrinsic differences in camera characteristics, such as placement and mobility. The static camera benefits more from higher resolution to capture small, distant objects, while the mobile camera benefits more from higher frame rates to adapt to rapid scene changes. These results highlight the importance of camera-specific configuration based on GPU budgets.

We leverage this observation to guide the design of our sampling configuration selection method. First, each camera conducts offline profiling to evaluate the accuracy of different frame rate and resolution combinations $(f, q)$ across different GPU budget levels. Given that retraining occurs within fixed-length retraining windows, which are further discretized into microwindows, the number of distinct GPU resource levels is limited. This results in a lookup table that maps GPU budgets to their corresponding optimal sampling configuration. 

At runtime, upon receiving its group’s GPU allocation $c_j$, each camera queries its table to select the optimal sampling configuration $(f^\star, q^\star)$. To balance data contributions across the $n_j$ group members, it scales the frame rate to $f^\star/n_j$ while keeping $q^\star$ unchanged. This ensures that the total volume of sampled frames aligns with the group’s compute capacity.

\subsubsection{\blue{Adaptive Weighted Bandwidth Allocation}}
\label{sec:bwallocation}
\blue{After sampling, each camera encodes frames based on available bandwidth and transmits them to the server. Since they all send to the same destination, cameras naturally share one or more network links, leading to contention and bandwidth bottlenecks. A naïve equal-allocation strategy can result in inefficient bandwidth usage, especially considering GPU resources can be allocated unevenly across cameras. We argue that ``utility-based'' bandwidth allocation, where each camera's bandwidth share is proportional to its GPU allocation, better aligns data delivery with compute capacity and leads to improved retraining quality.}

To illustrate this, we conduct a simple case study using cameras A and B in Fig.~\ref{fig:moti_conf}. Camera A starts retraining with a model accuracy of $28\%$ mAP, and camera B with $16\%$. To help B catch up and improve overall performance, we allocate $30\%$ of the GPU to A and $70\%$ to B. Each camera uses its optimal sampling configuration under the assigned GPU budget. We fix the total uplink bandwidth to $3\,Mbps$ and compare two bandwidth allocation strategies: equal allocation vs. GPU-proportional allocation (i.e., $0.9\,Mbps$ for A and $2.1\,Mbps$ for B).

\begin{table}[]
\caption{\blue{Retraining accuracy under equal vs. GPU-proportional bandwidth allocation.}}
\vspace{-1.2em}
\resizebox{\columnwidth}{!}{%
\begin{tabular}{@{}cccc@{}}
\hline
\hline
\begin{tabular}[c]{@{}c@{}} BW allocation\\schemes\end{tabular}  & \begin{tabular}[c]{@{}c@{}}Camera A\\mAP (\%)\end{tabular} & \begin{tabular}[c]{@{}c@{}}Camera B\\mAP (\%)\end{tabular}  & \begin{tabular}[c]{@{}c@{}}Overall \\mAP (\%)\\\end{tabular} \\ 
\hline
Equal (1.5 Mbps each) & 34.7  & 26.1 & 30.4   \\ 
Proportional (0.9/2.1) & 33.4   & 30.8    & 32.1  \\
\hline
\hline
\end{tabular}}
\label{tab:moti}
\vspace{-1.2em}
\end{table}
Table~\ref{tab:moti} shows that equal bandwidth allocation leads to lower retraining accuracy for the high-GPU camera B, likely due to delayed, dropped, or degraded frames. This results in  under-utilization of compute resources and reduced overall retraining accuracy. In contrast, proportional allocation allows each camera to deliver training frames in time to match its GPU share, thus improving retraining accuracy.

\blue{Achieving such GPU-proportional weighted bandwidth allocation is however challenging because of the distributed and heterogeneous nature of the network. Cameras connect to the server through diverse paths with varying conditions, and two types of constraints can occur together: (i) multiple cameras may share an uplink bottleneck with unknown capacity; and (ii) individual cameras, especially mobile ones, may also be constrained by their own weak local links.}

To address this, we design a distributed rate control mechanism based on the GAIMD (Generalized Additive Increase Multiplicative Decrease) congestion control algorithm~\cite{gaimd}. The core idea is to scale each camera’s aggressiveness in bandwidth competition according to its GPU share. By customizing the GAIMD parameters, each camera converges to a steady-state sending rate that approximates GPU-proportional bandwidth sharing.  

Specifically, each camera tunes its GAIMD parameters, the additive increase factor $\alpha$ and multiplicative decrease factor $\beta$, to control its transmission rate. This design leverages a known result: the steady-state throughput of a GAIMD flow is roughly proportional to $\alpha / (1 - \beta)$~\cite{gaimd-theory}. Based on this, upon receiving the GPU share weight $p_j$ from the server, each camera fixes $\beta = 0.5$ and sets $\alpha = p_j / n_j$, where $n_j$ is the number of cameras in its group $j$. \blue{The resulting GAIMD rate serves as the target sending rate for the video stream. The video encoder tracks this target and adjusts compression (quantization) in real time to stay close to it, while keeping the selected frame rate and resolution fixed. }

A key advantage of this approach is that it requires neither explicit coordination between cameras nor knowledge of the network topology. It automatically adapts to both shared bottlenecks and individual camera constraints. While some cameras may fall short of their target bandwidth due to local network conditions, the system still approximates GPU-proportional bandwidth allocation in a best-effort manner.

\vspace{-0.5em}
\subsection{Dynamic Camera Grouping}
\label{sec:grouping}
\blue{While the previous two subsections focus on steady-state operations for group retraining, \sysname must also handle the creation and update of camera groups to make retraining effective. This includes initial grouping to accommodate new retraining requests and periodic checks to determine whether existing groups remain valid or require regrouping. To keep the process lightweight, \sysname exploits metadata for fast pre-filtering, which narrows down candidate groups and avoids unnecessary computation. Final grouping and regrouping decisions are then based on actual accuracy improvements that directly reflect the goal of retraining. Algorithm~\ref{alg:grouping} outlines this process.}

\begin{algorithm}[t]
\caption{Dynamic Camera Grouping Algorithm} \label{alg:grouping}
\begin{algorithmic}[1] 
\State \textbf{Input:} retraining jobs \( \mathcal{J} \), new retraining request $\hat{r}$
\Procedure{GroupRequest}{$\mathcal{J}$, $\hat{r}$}
    \For{\( j \) in \( \mathcal{J} \)}
            \If{ $\forall$ $r$ in $j$, \( \left| r.t - \hat{r}.t \right| \leq \varepsilon \) and \( \left| r.loc - \hat{r}.loc \right| \leq \delta \)} \Comment{\textbf{Correlation filtering}}
            \State \( acc_j \gets j.\text{EVAL}(\hat{r}.subsamples) \)
            \If{$acc_j \geq \hat{r}.acc$} \Comment{\textbf{Performance check}}
            \State \( JobCandidate[j] \gets acc_j \)
            \EndIf
            \EndIf
            \EndFor
    \If{$JobCandidate \neq \emptyset$}
        \State \( \hat{j} \gets \text{argmax}(JobCandidate) \); $\hat{j} \gets \hat{j} \cup \{\hat{r}\}$
    \Else
        \State  $\hat{j} \gets InitializeNewJob(\hat{r})$; $\mathcal{J} \gets \mathcal{J} \cup \{\hat{j}\}$
    \EndIf
\EndProcedure
\Procedure{UpdateGrouping}{$\mathcal{J}$}
\While{each retraining window $n$ ends}
    \For{\( j \) in \( \mathcal{J} \)}
    \For{\( r \) in \( j \)}
         \State \( r.acc_n \gets j.\text{EVAL}(r.subsamples) \)
            \If{$\frac{r.acc_{n}-r.acc_{n-1}}{r.acc_{n-1}}<-p$ $(p>0)$} 
            \State $j.Remove(r)$; $ $ $r.Update(t, loc)$; 
            \State GroupRequest($\mathcal{J}$, $r$)
            \EndIf
            \EndFor
            \EndFor
            \EndWhile
\EndProcedure
\end{algorithmic}
\end{algorithm}

\myparagraph{Grouping initialization for new retraining requests:} \blue{At runtime, each camera detects data drift locally and issues a retraining request when drift is observed. Several existing techniques \cite{foot1, foot2, foot3} can be used for drift detection.} The request includes metadata (request time and location), sampled frames, and a copy of the device’s lightweight model. Upon receiving the request, \sysname checks whether any ongoing retraining jobs (groups) show temporal and spatial correlations with the new request. It compares the new request’s metadata against those in ongoing jobs to determine if they fall within a predefined time window and geographical range (Line 4). The rationale is that cameras operating close in space often experience similar data drifts at similar times. If no correlated group is found, \sysname initiates a new retraining job for the request, starting with the device's sent model and sampled frames (Line 11).

If correlated groups are found, \sysname evaluates these jobs' model performance using a small subset of the sample frames from the new request (Line 5). The model that shows the most significant improvement on the new request is selected, and the new request is then integrated into the corresponding job; this involves adding its metadata to the job's metadata and aggregating its sample frames into the job's training data (Line 9). This performance check ensures that the grouping decision is grounded in actual model accuracy rather than relying solely on metadata similarity, thereby preventing incorrect group assignments that could harm retraining accuracy. Since this grouping process is efficient, the delay it introduces is negligible compared to the duration of a retraining window.

\myparagraph{Periodic reevaluation of existing groups:} 
Since video content evolves over time, retraining groups must be updated dynamically. For example, mobile cameras in a fleet may initially share similar scenes and be grouped together, but changes in routes can cause their data distributions to diverge. To account for such dynamics, \sysname periodically reassesses existing groups at the end of each retraining window. It iteratively evaluates the performance of every group model on each camera member and compares it to the performance in the previous window (Line 16-17). \blue{If a camera’s accuracy drops beyond a threshold, it signals that the camera has undergone a second drift during the window and no longer aligns with the group's data distribution. The camera is then removed from the group. There can be situations where only some cameras are removed from a group, and others where all are removed due to group-wide new drift. In either case, each removed camera is treated as a ``new'' retraining request with updated metadata and reprocessed through the initial grouping logic. This allows removed cameras to form new groups if they now share similar data distributions.}

\vspace{-0.5em}
\section{{Implementation and Experimental Setup}}
\label{sec:implementation}
We implement \sysname in Python, using the Ultralytics framework \cite{ultralytics} (built on PyTorch \cite{pytorch}) for both model training and inference. The system runs on a server equipped with five NVIDIA GeForce RTX 4090 GPUs. To emulate cameras and their transmission to the server, we use a hybrid setup that combines Docker-based container emulation and NS-3~\cite{ns-3} simulation.

\blue{For each retraining window, we run an NS-3 simulation to generate per-camera bandwidth traces over time, based on GAIMD-driven, GPU-aware bandwidth allocation (\S\ref{sec:bwallocation}). Each camera is emulated as an independent Docker container on the server. At the start of each window, each container encodes its video at a selected frame rate and resolution (\S\ref{sec:sampling}) using FFmpeg. During streaming, video is split into 1-second segments. For each segment, FFmpeg’s target bitrate is set to the average bandwidth of the corresponding NS-3 trace segment, so that compression adapts automatically. Linux traffic control (tc) enforces the NS-3 trace by shaping outgoing packets to match the simulated bandwidth.}

\myparagraph{Datasets:} We evaluate \sysname using three diverse  datasets. (i)\textit{ CityFlow dataset \cite{cityflow}} includes over 3 hours of synchronized videos captured by $40$ traffic cameras at $10$ city intersections. (ii) \textit{MDOT dataset \cite{mdot}} consists of a total of $155$ groups of video clips (totaling more than $259k$ frames) captured by five drones. We arranged video clips from the same camera chronologically to create continuous video streams. (iii) \textit{CARLA simulated datasets}: The core concept behind \sysname involves leveraging the correlations between multiple cameras. To assess how these correlations affect \sysname's performance, it is essential to have datasets that feature videos with varying degrees of similarity. However, such datasets are rare in public collections. To address this, we employ the CARLA simulator \cite{carla} to create multiple datasets. CARLA is an autonomous driving simulator built on the Unreal Engine \cite{unrealengine}, which has been used in developing industrial autonomous vehicle systems like Apollo \cite{apollo} and Autoware \cite{autoware}. We generate traffic flows, place traffic cameras at varying distances from each other, and record the videos. By altering the positions of the cameras—from close proximity to widely spaced—we generate datasets that exhibit different levels of camera similarity.

\myparagraph{Models:} We demonstrate \sysname’s performance on two machine learning tasks – object detection and instance segmentation. Instance segmentation is a more complex computer vision task compared to detection. For object detection, we use YOLO11-Nano and YOLO11-X \cite{yolo11} as student and teacher models, respectively. Instance segmentation used YOLO11n-Seg and YOLO11x-Seg for the student and teacher models. All models are pre-trained on COCO \cite{coco} datasets. 

\myparagraph{Baselines:} We compare \sysname with the following continuous learning frameworks:

\setlist[itemize]{topsep=0pt, partopsep=0pt, itemsep=0pt, leftmargin=8pt, parsep=0pt}
\begin{itemize}
\item \myparagraph{\textit{Naive baseline:}} This baseline retrains a separate model for each camera (no grouping) and allocates GPU resources uniformly across all concurrent retraining jobs (without optimizing GPU usage). Each camera uses a fixed sampling configuration, and bandwidth is evenly shared among cameras (without optimizing bandwidth efficiency).

\item \myparagraph{\textit{Ekya:}} Ekya \cite{ekya} supports both inference and retraining tasks on edge devices by carefully scheduling GPU resources between these jobs. Unlike Ekya, \sysname allocates the server's GPU resources solely for retraining jobs, as inference is handled on edge devices. For a fair comparison, we evaluate Ekya in a retraining-only setting. Although Ekya utilizes more advanced GPU allocation mechanisms than the naive baseline, it retrains separate models for each camera (independent retraining) and does not exploit potential similarities in data drift across cameras for grouped retraining, as \sysname does.

\begin{figure*}[t]
    \newcolumntype{C}{>{\centering\arraybackslash}X}
    \centering
    \begin{minipage}{0.6\linewidth}
    \includegraphics[width=\textwidth]{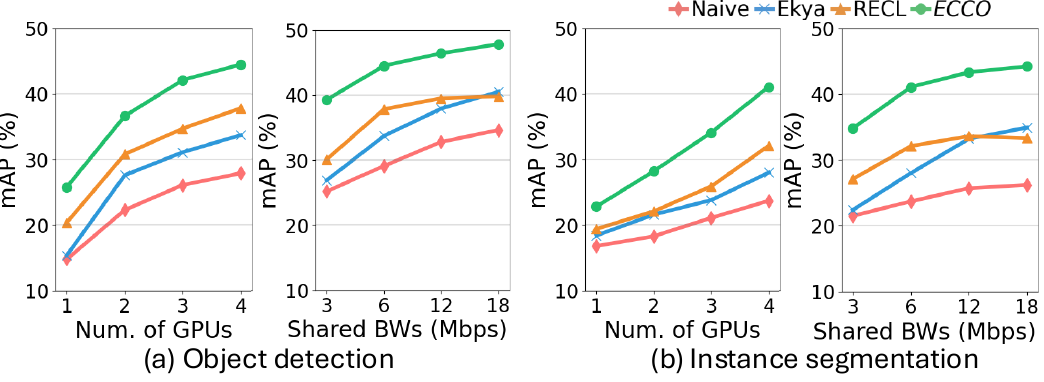}
    \vspace{-2.2em}
    \caption{\blue{Average accuracy across different GPU resources and shared bandwidth resources.}}
    \label{fig:e2e}
\end{minipage}
\hfill
\begin{minipage}{0.38\linewidth}
    \includegraphics[width=\textwidth]{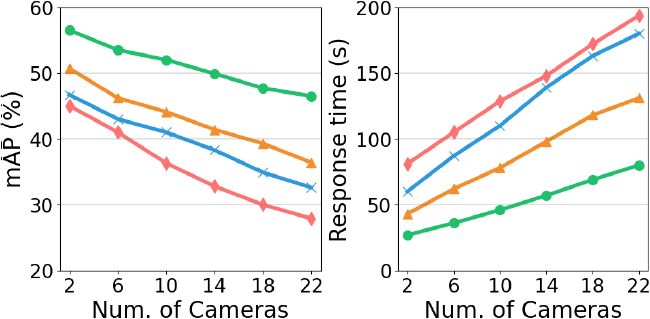}
    \vspace{-2em}
    \caption{Average retraining accuracy and response time scaling across multiple cameras.}
    \label{fig:scale}
\end{minipage}
\vspace{-1em}
\end{figure*}
\item \myparagraph{\textit{RECL:}} RECL \cite{recl} selects a historical model from a shared ``model zoo'' to initialize retraining, aiming to accelerate convergence by starting from a previously trained model. While this strategy improves over Ekya, RECL still performs retraining independently for each camera. As the official implementation is unavailable, we re-implemente RECL’s GPU allocation algorithm following the descriptions in the paper. To build the model zoo, we fine-tune a student model on the first two minutes of video from each camera and store each resulting model separately. RECL also adopts an adaptive frame uploading mechanism from AMS \cite{ams}, which adjusts each camera’s sampling frame rate based on scene dynamics. However, this adaptation is driven purely by video content and does not align the sampling configuration with GPU allocation, nor does it coordinate bandwidth allocation. Since RECL demonstrates substantial improvement over AMS in its evaluation, we do not compare \sysname against AMS.
\end{itemize}

\myparagraph{Metrics:} We evaluate \sysname and the baselines along two dimensions: (i) Inference accuracy: We use mean Average Precision (mAP), a standard metric for both object detection and instance segmentation tasks \cite{voc,coco}. mAP quantifies both precision and recall across various Intersection over Union (IoU) thresholds,  reflecting the model's accuracy in predicting bounding boxes (detection output) and segmentation masks (segmentation output). (ii) Response time: The duration required to retrain models to achieve a predefined accuracy after retraining is triggered. 
\section{Evaluation}
\label{sec:evaluation}
We evaluate \sysname on two video analytics tasks using three datasets. Our evaluation focuses on the following aspects:

\S\ref{sec:exp_end2end} What is the end-to-end performance of \sysname compared to leading baselines?

\S\ref{sec:exp_scale} How well does \sysname scale with increasing workloads?

\S\ref{sec:exp_similarity} How does camera similarity impact \sysname's performance advantage over baseline methods?

\S\ref{sec:exp_modules} How effective are the three modules of \sysname?

\S\ref{sec:exp_benefit} How does \sysname improve responsiveness to retraining requests, especially under low-bandwidth conditions?

Our key findings include:
\setlist[itemize]{topsep=0pt, partopsep=0pt, itemsep=0pt, leftmargin=8pt, parsep=0pt}
\begin{itemize}
\item \blue{With the same compute and communication resources, \sysname improves mAP by $6.7\%$-$16.6\%$ for object detection and $9.3\%$-$18.1\%$ for instance segmentation, consistently outperforming all baselines.}
\item \sysname can scale to support $3.3\times$ more cameras than baselines while maintaining the same accuracy.
\item \sysname's three modules jointly enable dynamic and accurate camera grouping, improve GPU and bandwidth efficiency, and enhance retraining quality.
\item \sysname reduces response times by more than $5\times$ compared to baselines under low-bandwidth conditions. These benefits stem from group retraining’s data aggregation and natural model reuse within a group.
\end{itemize}

\vspace{-1em}
\subsection{End-to-End Evaluation}
\label{sec:exp_end2end}
We first evaluate the end-to-end performance of \sysname against three baselines under varying GPU and bandwidth constraints. The experiments use a fixed workload consisting of all 6 cameras from Scene 03 of the CityFlow dataset. For the Naive baseline and Ekya, each camera samples frames at $5\,FPS$ with a vertical resolution of $960$, \blue{reflecting a default high-rate, high-resolution setting}. RECL adjusts only the frame rate. As none of the baselines consider shared or local bandwidth constraints, we assume only a shared bandwidth constraint in this experiment and equal bandwidth sharing across cameras for all baselines. We consider two experimental conditions: (i) varying the number of GPUs while fixing total shared bandwidth at $6\,Mbps$ \blue{(i.e., corresponding to $1\,Mbps$ per camera on average, used as a representative constrained bandwidth setting)}, and (ii) varying the total shared bandwidth while fixing the number of GPUs at 4. \blue{We use 4 GPUs due to testbed limits, emulating a small slice of a larger deployment.} We report the average accuracy across all cameras in Fig.~\ref{fig:e2e}.

\blue{Overall, \sysname consistently outperforms all baselines. Under the same compute and communication budgets, it improves mAP by $6.7\%$-$16.6\%$ for object detection and $9.3\%$-$18.1\%$ for instance segmentation over the baselines.} For object detection, \sysname requires only $1.8\times-2.4\times$ fewer GPUs to maintain an mAP of $35\%$, due to its group retraining strategy. By merging retraining across correlated cameras, \sysname reduces the number of concurrent models, enabling more GPU resources per model on average. Such efficiency is further improved by optimizing GPU resource allocation across groups. \sysname also achieves the same detection accuracy (around $40\%$ mAP) while using only $25\%$–$33\%$ of the bandwidth required by the baselines. This benefit stems from its transmission control module, which jointly adjusts sampling configurations and bandwidth sharing to match GPU allocation. This design ensures that available bandwidth is used fully and efficiently.

Among the baselines, RECL performs best due to its model reuse and GPU allocation strategy. However, as it reduces the sampling rate based on scene dynamics and does not adapt to bandwidth availability, it misses the opportunity to benefit from more training data when more bandwidth is available. In addition, RECL introduces overhead from continuously updating a large model zoo and retraining a model selector, which is not reported here. Finally, there is no guarantee that a historical model will perfectly match the current data drifts. In contrast, \sysname directly identifies and leverages current correlations among retraining requests without incurring much extra costs.

\begin{figure}[]
\centering
\captionsetup{font={small, stretch=0.8}}
    \includegraphics[width=\linewidth]{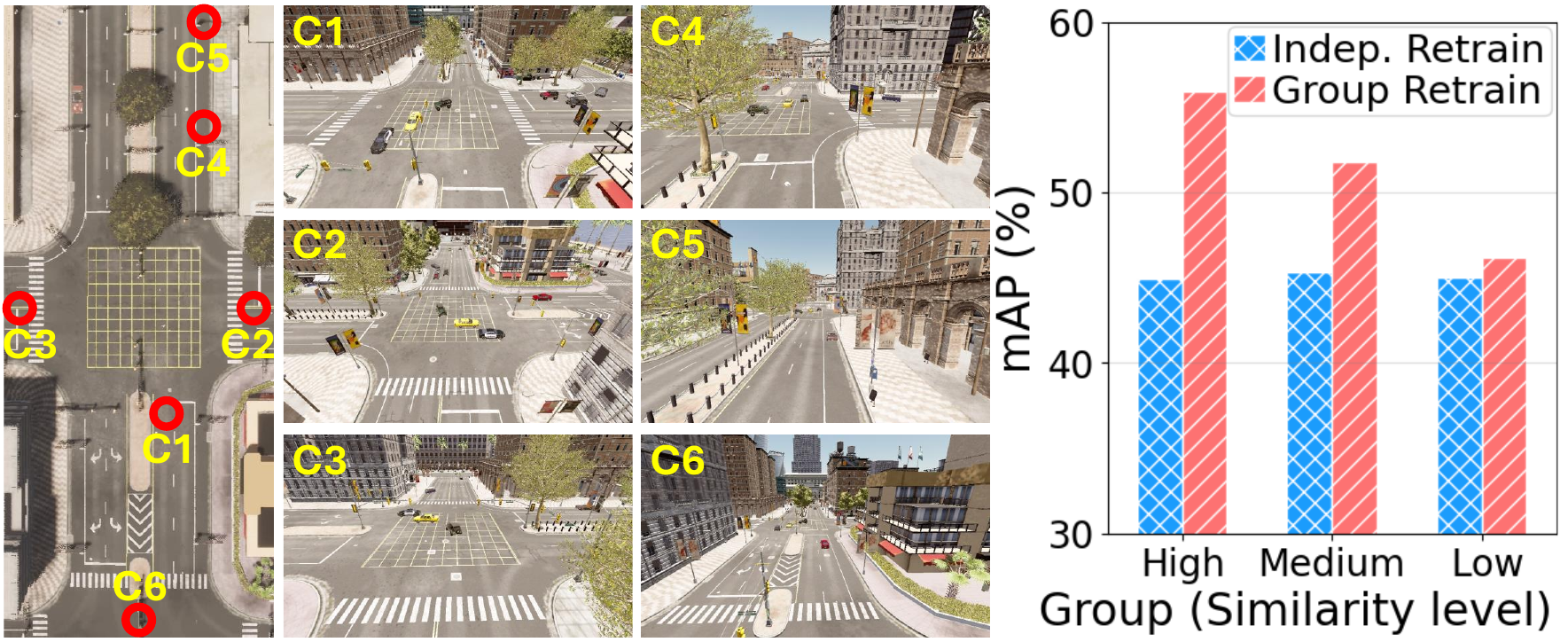}
        \vspace{-2em}
    \caption{\blue{Impact of camera similarity. Left: Camera placement and example frames from six cameras in a CARLA town scene. Cameras are manually grouped into high (C1–C2–C3), medium (C1–C4–C5), and low (C1–C5–C6) similarity groups. Right: Retraining accuracy under independent vs. group retraining across similarity levels.}}
    \vspace{-1em}
    \label{fig:similarity}
\end{figure}
\vspace{-0.5em}
\subsection{Scalability of \sysname}
\label{sec:exp_scale}
We evaluate the scalability of \sysname by testing it on workloads with an increasing number of cameras. Specifically, we use video streams from up to 22 cameras in Town 3 of the CARLA simulator, and the detailed scene and camera placements are visualized in the appendix. The experiment is conducted on the object detection task using a fixed compute budget of 4 GPUs and a simulated shared bandwidth of $50\,Mbps$. Fig.~\ref{fig:scale} reports the average retraining accuracy and response time (using a mAP threshold of 0.4) across all cameras.

As workload increases, the retraining accuracy of all three baselines drops significantly, and their response times grow quickly. This degradation stems from their independent retraining approach, which causes the compute demand to grow linearly with the number of cameras. In contrast, \sysname exhibits a more moderate degradation due to its group retraining strategy. Compared to the best-performing baseline, RECL, \sysname supports $3.3\times$ more cameras using the same resource budget to achieve a similar mAP of $46\%$. In terms of responsiveness, under the 22-camera workload, \sysname reduces the response time to $41.3\%$–$61.1\%$ of the baselines.

\vspace{-0.5em}
\subsection{Impact of Camera Similarity}
\label{sec:exp_similarity}
This section provides an intuitive understanding of when \sysname's group retraining is beneficial compared to independent retraining, and when it is not. We evaluate how the effectiveness of group retraining depends on the similarity among cameras within a group. To visualize and control similarity, we simulate six cameras in a region of Town 10 in the CARLA simulator and show their positions and fields of view in Fig.\ref{fig:similarity}. We disable \sysname's grouping module and manually construct three groups of three cameras each: a high-similarity group (C1–C2–C3), where cameras capture highly overlapping scenes with similar content; a medium-similarity group (C1–C4–C5), where cameras are in nearby locations with partially correlated views; and a low-similarity group (C1–C5–C6), where cameras observe distinct and non-overlapping scenes. We simulate a sudden rain event (weather-induced data drift) and apply group retraining to each group using a fixed 3-GPU and $3\,Mbps$ shared bandwidth budget. We compare the results to independent retraining using Ekya, configured with the same resources.

\begin{figure}[]
\centering
\captionsetup{font={small, stretch=0.8}}
    \begin{subfigure}[b]{0.98\linewidth}
        \includegraphics[width=\textwidth]{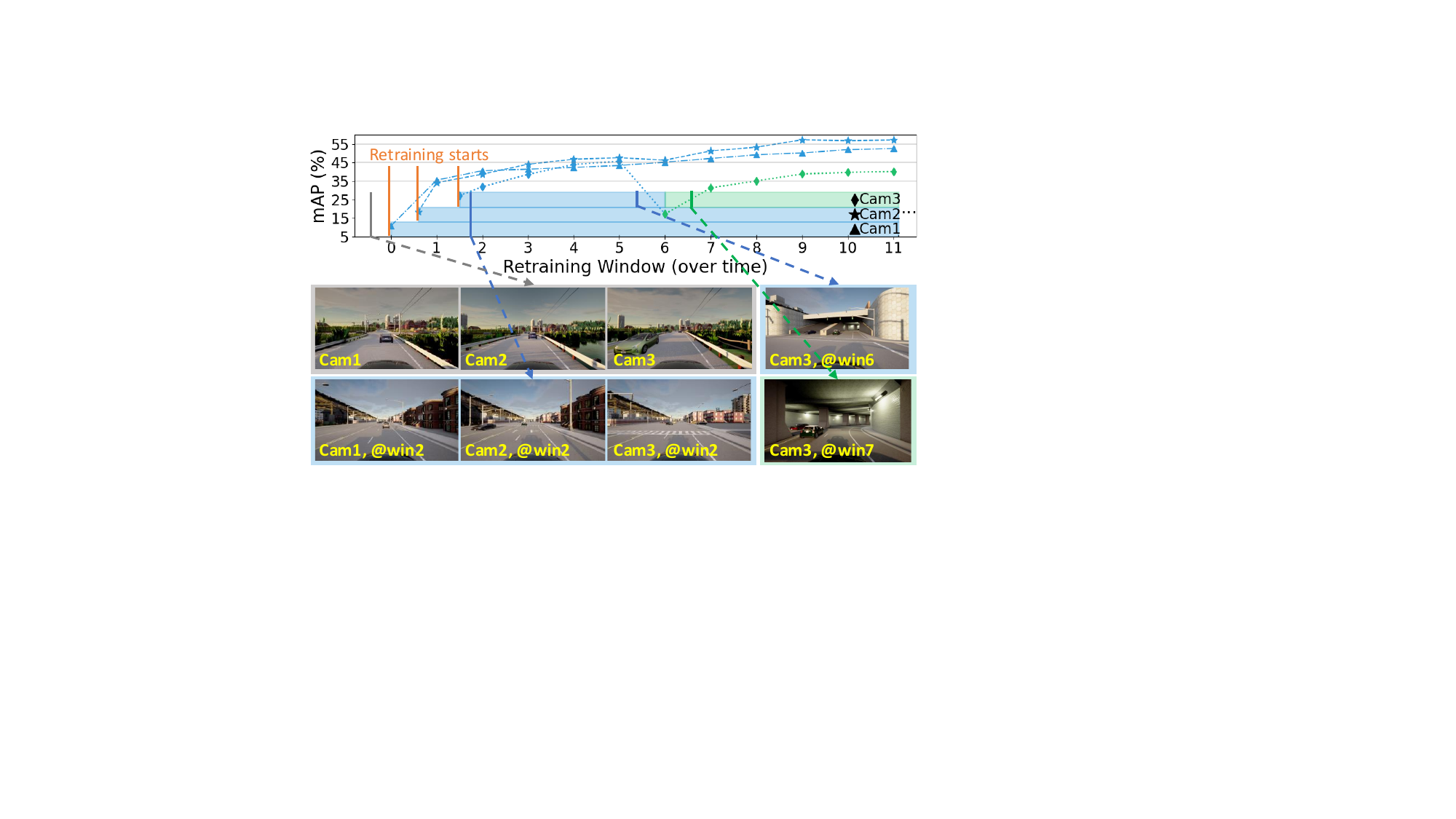}
        \vspace{-2em}
    \end{subfigure}
    \caption{An example of \sysname's dynamic camera grouping. Lines show the retraining accuracy over time for each camera, differentiated by marker types. Bars below represent camera groupings, with identical colors indicating the same group. \sysname dynamically regroups camera 3 when it no longer benefit from the current group due to scene divergence.}
    \vspace{-1em}
    \label{fig:grouping}
\end{figure}
As shown in Fig.\ref{fig:similarity}, group retraining outperforms independent retraining when camera similarity is high, improving mAP by up to $11.9\%$. However, this advantage diminishes with decreasing similarity, and in the low-similarity group, group retraining provides little improvement due to limited shared feature across cameras.

\vspace{-0.5em}
\subsection{Performance of Modules in \sysname}
\label{sec:exp_modules} 
In this subsection, we evaluate the effectiveness of the three modules in \sysname, respectively.

\subsubsection{\myparagraph{Dynamic camera grouping}}
\label{sec:exp_grouping}
To evaluate the effectiveness of \sysname's dynamic camera grouping, we collect driving videos from three vehicles navigating a town map in CARLA. Mobile cameras introduce a more challenging scenario due to rapid and frequent scene changes, which require dynamic assessment and regrouping. Fig.~\ref{fig:grouping} shows an example of dynamic grouping in \sysname, including each camera's grouping status and retraining accuracy over time. We also present video frames at selected timestamps to qualitatively verify the grouping behavior.

As the vehicles move sequentially from suburban to urban areas, their cameras experience similar data drift caused by background transitions, leading to a drop in model accuracy to below 15\% mAP for all three cameras. \sysname first receives a retraining request from camera 1 and starts a new retraining job. Later, it adds cameras 2 and 3 to the same job upon receiving their requests, as they share similar metadata with camera 1, and the ongoing training model from camera 1 performs better on their data than their current local models.

As the shared model is retrained, all three cameras benefit from improved accuracy. However, during retraining window 6, camera 3 takes a different route and enters a tunnel, while cameras 1 and 2 continue along the city road. This causes camera 3’s data distribution to diverge significantly. When \sysname reassesses the group model's performance on camera 3 at the end of the window, it detects a significant accuracy drop, indicating camera 3 no longer aligns with the current group. \sysname then removes camera 3 from the group and treats its case as a new retraining request. Since no existing group matches its updated metadata, a separate retraining job is initiated for camera 3.

\begin{figure}[t]
\centering
\captionsetup{font={small, stretch=0.8}}
    \begin{subfigure}[b]{0.49\linewidth}
        \vspace{-0em}
        \includegraphics[width=\textwidth]{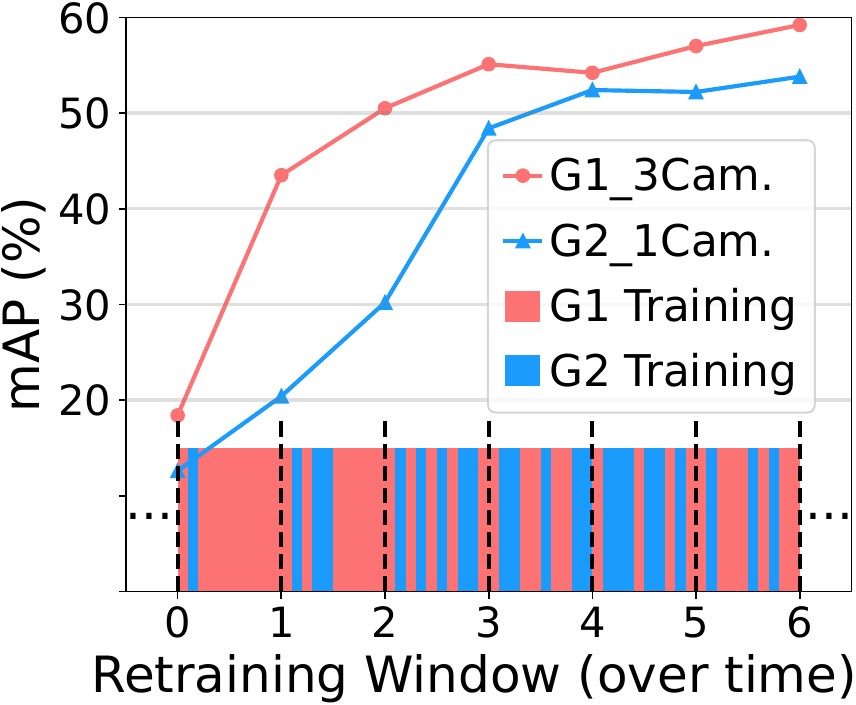}
        \caption{RECL's GPU allocator}
        \vspace{-0.8em}
    \end{subfigure}
    \hfill
    \begin{subfigure}[b]{0.49\linewidth}
        \vspace{-0em}
        \includegraphics[width=\textwidth]{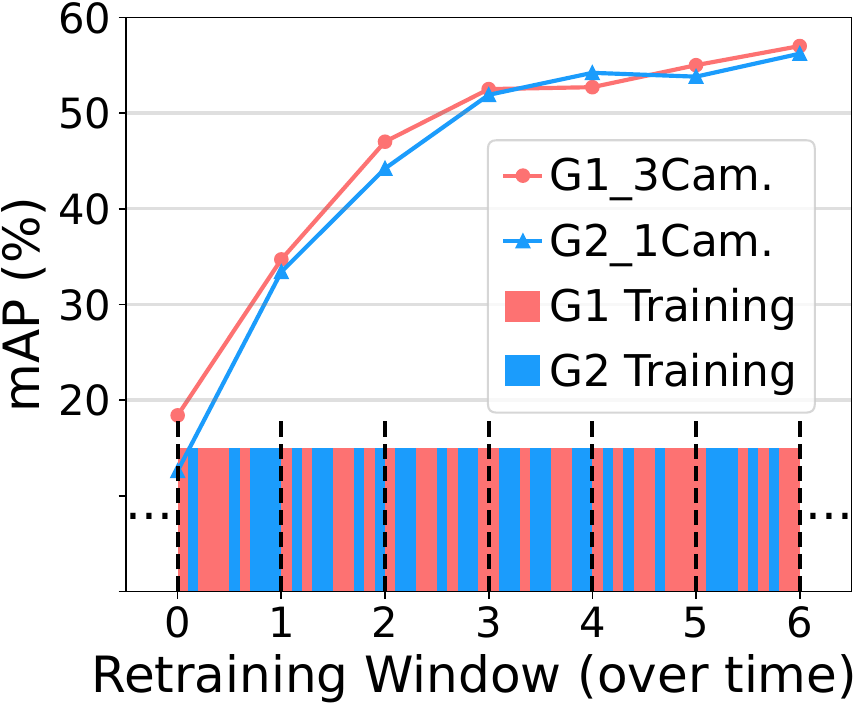}
        \caption{\sysname's GPU allocator           }
        \vspace{-0.8em}
    \end{subfigure}
    \caption{\sysname with RECL's GPU allocator vs. its native allocator. The lines above show average accuracy for two distinct groups (3 cameras vs. 1 camera). The bars below represent GPU allocation over time. RECL's allocator focuses on overall accuracy, neglecting smaller groups, while \sysname balances overall accuracy and fairness among cameras.} 
    \vspace{-1.19em}
    \label{fig:schedule}
\end{figure}
\subsubsection{\myparagraph{GPU allocator}}
\label{sec:exp_scheduler}
We evaluate the performance of \sysname's GPU allocator by replacing it with RECL's allocator. The experiment used four drone videos from the MDOT dataset, featuring three drones operating in an adjacent area and one drone in a distinct area. This setup ultimately divided the drones into two groups: one with three cameras and another with a single camera. Fig.~\ref{fig:schedule} illustrates the GPU resource allocation for both allocators and the average retraining accuracy over time for each group. As both allocators share GPU resources on a time-shared basis, we used a ``one-hot bar'' to display the GPU allocation results.

When applying the RECL allocator to our system, it allocated most of the GPU resources to group 1 in the first two retraining windows.
As a result, group 2 experienced resource starvation, leading to a significant accuracy gap of up to $23\%$ mAP between the two groups. This is because the RECL allocator is designed to maximize the total system accuracy, favoring the group with more cameras due to their larger impact on overall accuracy improvement. In contrast, \sysname offers a more balanced approach. It achieves a near-synchronous accuracy increase among different groups. This is because \sysname's allocator not only seeks to improve overall accuracy, but it also considers  fairness by boosting the low-performance groups.

\begin{figure}[t]
\centering
\captionsetup{font={small, stretch=0.8}}
    \begin{subfigure}[b]{0.325\linewidth}
        \includegraphics[width=\textwidth]{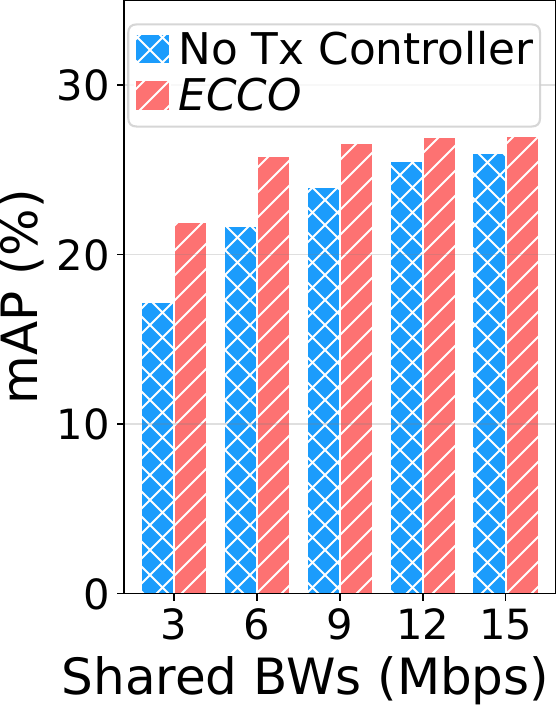}
        \vspace{-1.5em}
    \end{subfigure}
    \hfill
    \begin{subfigure}[b]{0.65\linewidth}
        \includegraphics[width=\textwidth]{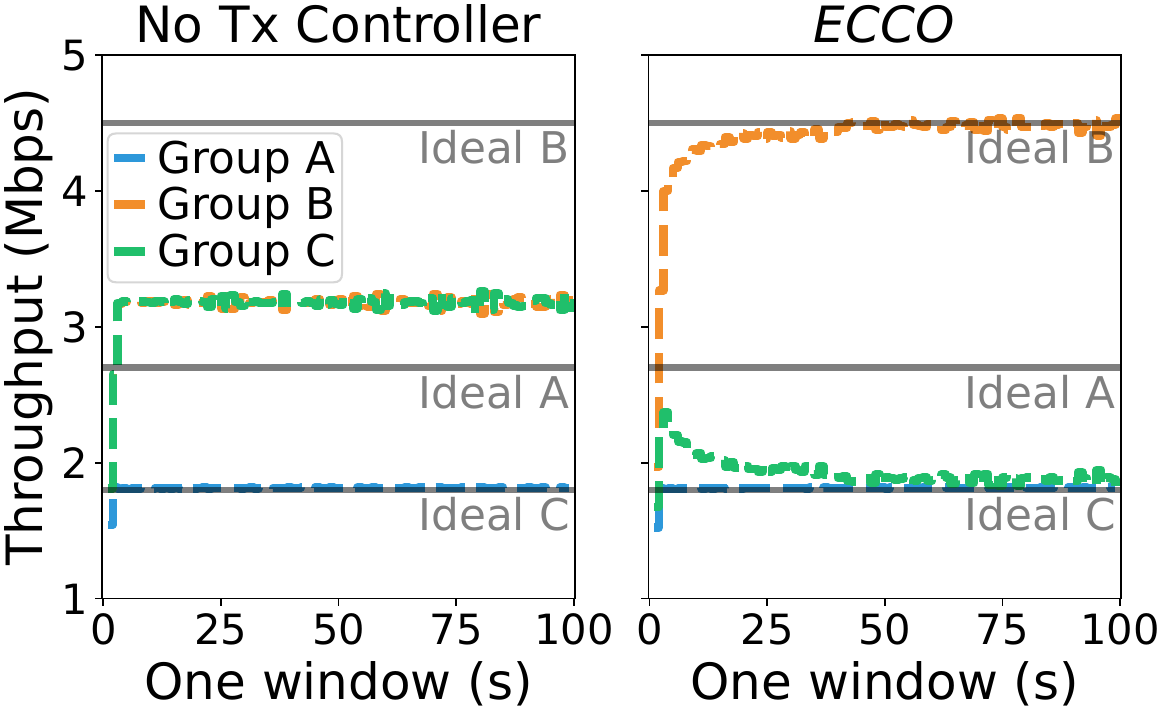}
        \vspace{-1.5em}
    \end{subfigure}
    \vspace{-0.5em}
    \caption{Ablation study of the transmission controller. Left: Retraining accuracy under varying shared bandwidths, showing that the controller improves accuracy, especially under limited bandwidth. Right: Per-group bandwidth traces at $9\,Mbps$ shared bandwidth. The controller approximates GPU-proportional bandwidth allocation, whereas the baseline deviates significantly from the ideal allocation target.} 
    \vspace{-1em}
    \label{fig:bandwidth}
\end{figure}
\subsubsection{\myparagraph{\blue{Resource-Aware Transmission Controller}}}
\label{sec:exp_uploadrate}
\blue{We evaluate our transmission controller through an ablation study. In the ablated baseline, the controller is disabled at each camera: all cameras sample frames at a fixed rate of $5\,fps$ and resolution of $960$. Bandwidth sharing follows the traditional AIMD rule ($\alpha=1$, $\beta=0.5$) across cameras, meaning each camera competes equally for shared bandwidth, subject to its own local uplink cap. All other system components remain unchanged. We use 6 cameras from the CARLA dataset, evenly grouped into three groups (A, B, C), and fix the GPU budget to 1 GPU. The total shared bandwidth is varied from $3\,Mbps$ to $15\,Mbps$. To emulate heterogeneous network conditions, we cap the uplink of the two cameras in Group A to $1\,Mbps$.}

Fig.~\ref{fig:bandwidth} (Left) shows the average retraining accuracy across all cameras. As the shared bandwidth increases, the performance bottleneck shifts from communication to computation, leading to an increase in accuracy for both methods that eventually plateaus. \sysname reaches its peak accuracy using only one-third of the bandwidth required by the baseline and achieves up to $4.7\%$ higher accuracy under limited bandwidth ($3\,Mbps$). Fig.~\ref{fig:bandwidth} (Right) zooms in on a retraining window at $9\,Mbps$ shared bandwidth, where GPU allocation across Groups A, B, and C is approximately in a 3:5:2 ratio. It compares the per-group bandwidth traces of the two methods against the ideal GPU-proportional target. \sysname closely approximates the target allocation, with Groups B and C proportionally sharing the remaining bandwidth after Group A’s local constraint is saturated. In contrast, the baseline deviates significantly due to the lack of rate differentiation. These results demonstrate that aligning communication with compute resources improves retraining accuracy, and that our transmission controller achieves compute-aware, adaptive bandwidth sharing under heterogeneous network conditions.
\vspace{-1em}
\subsection{\sysname's Benefits in Responsiveness}
\label{sec:exp_benefit}
In \S\ref{sec:exp_scale}, we showed that \sysname improves responsiveness to data drift through optimized compute and communication resource usage. Beyond resource efficiency, here we highlight two additional factors that contribute to its responsiveness: natural model reuse and data aggregation within a group. To isolate and evaluate these effects, we conduct intra-group experiments using Group 1 in Fig.\ref{fig:schedule}, which includes three drone video streams. We compare \sysname with two methods: 
(i) RECL, which selects a historical model as the retraining starting point;
(ii) \sysname+RECL, which combines group retraining with historical model reuse.

\begin{figure}[]
\centering
\captionsetup{font={small, stretch=0.8}}
\begin{minipage}[t]{0.48\linewidth}
\centering
\includegraphics[width=\textwidth]{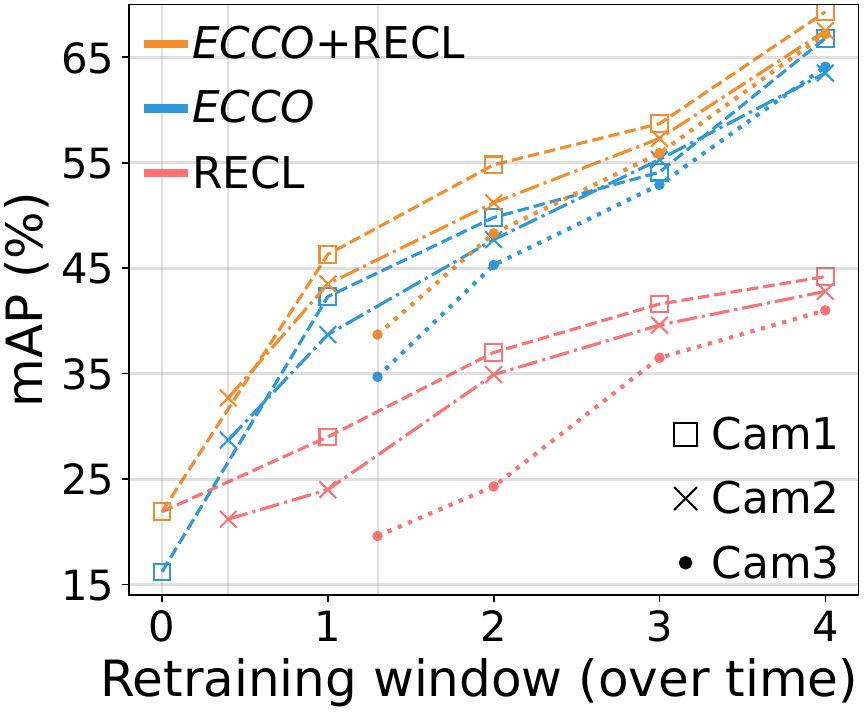}
\vfill
\vspace{-1em}
\caption{Retraining accuracy for each camera within a group over time. Group retraining enhances the initial accuracy of later cameras naturally.}
\vspace{-1.5em}
\label{fig:reuse}
\end{minipage}
\hfill
\begin{minipage}[t]{0.48\linewidth}
\centering
\includegraphics[width=\textwidth]{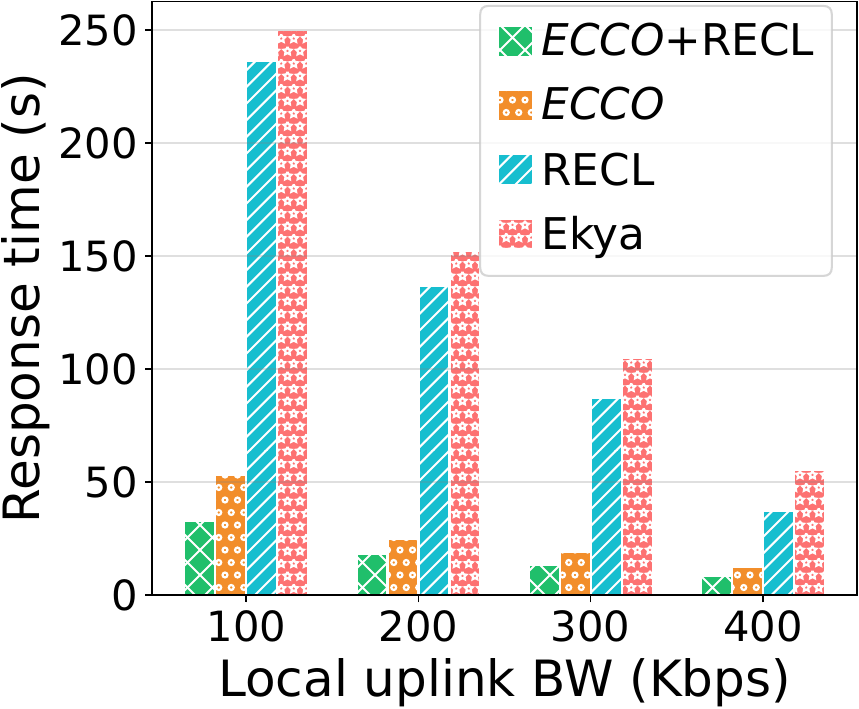}
\vfill
\vspace{-1em}
\caption{Average response time across cameras under low-bandwidth conditions. Group retraining enhances responsiveness through data aggregation.}
\vspace{-1.5em}
\label{fig:responsetime}
\end{minipage}
\end{figure}

\myparagraph{Natural model reuse.} Fig.~\ref{fig:reuse} shows each camera's retraining accuracy over time. For cameras 2 and 3, \sysname and \sysname+RECL achieve up to $15\%$ higher initial mAP than RECL. This is because group retraining allows later retraining requests to start from a model that has already been partially updated using data from earlier cameras in the same group. In contrast, RECL relies on static historical models, which may not perfectly match the current, drifted data distribution. For camera 1, RECL achieves higher initial accuracy because it reuses an appropriate historical model, whereas \sysname starts retraining from scratch. \sysname+RECL inherits the strengths of both approaches and consistently yields the highest initial accuracy across all cameras.

\myparagraph{Data aggregation.} Group retraining improves responsiveness under poor network conditions, which are common in mobile scenarios such as drones or vehicles. Fig.~\ref{fig:responsetime} shows the average time required to reach $35\%$ mAP under various low-bandwidth constraints on each camera's local uplink. RECL and Ekya exhibit up to $5\times$ longer response times compared to group retraining methods. This is because individual retraining must wait for sufficient data from a single camera, whereas group retraining aggregates data streams from multiple cameras, effectively increasing the available training data and speeding up training. Incorporating RECL into \sysname further reduces response time by initializing retraining from a stronger starting point.

\section{Related Work} 
\label{sec:related}
\myparagraph{Live video continuous learning:} Prior studies have focused on building video analytics systems to provide high accuracy, low cost, and fast responses.
These systems employ techniques such as model merging \cite{gemel}, model architecture pruning \cite{approxnet,fbnet}, model distillation \cite{noscope, real}, configuration adaptation \cite{madeye, chameleon, vetl, live}, and frame selection \cite{glimpse, reducto}. However, all these efforts have aimed to optimize only the \textit{inference} accuracy or the compute/network costs of DNN \textit{inference}. In contrast, our work focuses on serving \textit{continuous learning} for live video analytics, a relatively unexplored focus area until recently. One pioneering effort in this domain is Ekya \cite{ekya}, which introduces a scheduler that optimally allocates GPU resources between retraining and inference tasks on edge servers. Building on this, RECL \cite{recl} further integrates model reuse with continuous retraining to enhance resource efficiency and responsiveness to data drift in live videos. The most recent study, AdaInf \cite{adainf}, manages GPU resource allocation to ensure service level objectives (SLOs) are guaranteed across multiple retraining models.

However, these systems share a common limitation: they handle retraining requests from different cameras independently, neglecting the similarity and potential for synergy between them, which can result in redundant retraining costs. \sysname addresses this issue by recognizing and exploiting the potential correlations between different camera feeds. By merging similar retraining requests, we improve the resource efficiency of the system.

\myparagraph{Leveraging cross-camera correlations:} Cross-camera correlations have been well recognized and utilized in previous work. In the computer vision community, these correlations are extensively studied in two main tasks: person re-identification (re-id) and multi-target, multi-camera (MTMC) tracking. Many studies have proposed new neural network architectures that use multi-camera correlations to address these tasks \cite{ccc1, ccc2, ccc3, ccc4}. In the systems literature, Chameleon \cite{chameleon} and Spatula \cite{spatula} are two notable examples. Chameleon uses the temporal and spatial correlations among different videos to reduce the cost of neural network configuration profiling. Similarly, Spatula exploits cross-camera correlations to lower the inference costs in applications such as re-id and MTMC tracking.

\sysname stands apart from these approaches, as it aims to reduce the continuous learning costs of video analytics systems through cross-camera correlations. Guided by this objective, we present a new concept—group retraining. We have also developed an end-to-end framework that optimizes the use of compute and communication resources.

\section{Conclusion}
\label{sec:discussion}
In this paper, we introduced \sysname, a novel video analytics framework that significantly enhances the efficiency of continuous learning by leveraging cross-camera correlations. ECCO smartly groups cameras experiencing similar data drifts to retrain a shared model, thereby reducing redundancy and optimizing resource utilization in terms of both computing power and data transmission. Extensive evaluations on multiple datasets demonstrated that ECCO markedly outperforms existing systems in accuracy,  efficiency, and scalability. 

\bibliographystyle{ACM-Reference-Format}
\newpage
\bibliography{refs}
%
%
%
%
%




\appendix

\newpage
\section{Appendix}
Here we provide more details on the dataset generation using the CARLA simulator. We set up fixed traffic cameras at various locations in Town 3 within the CARLA ecosystem. These cameras are used for assessing the scalability of \sysname (Section \ref{sec:exp_scale}). The locations and orientations of these traffic cameras are illustrated in the figure below.
\begin{figure}[H]
\centering
\captionsetup{font={small, stretch=0.8}}
    \begin{subfigure}[b]{0.98\linewidth}
        \includegraphics[width=\textwidth]{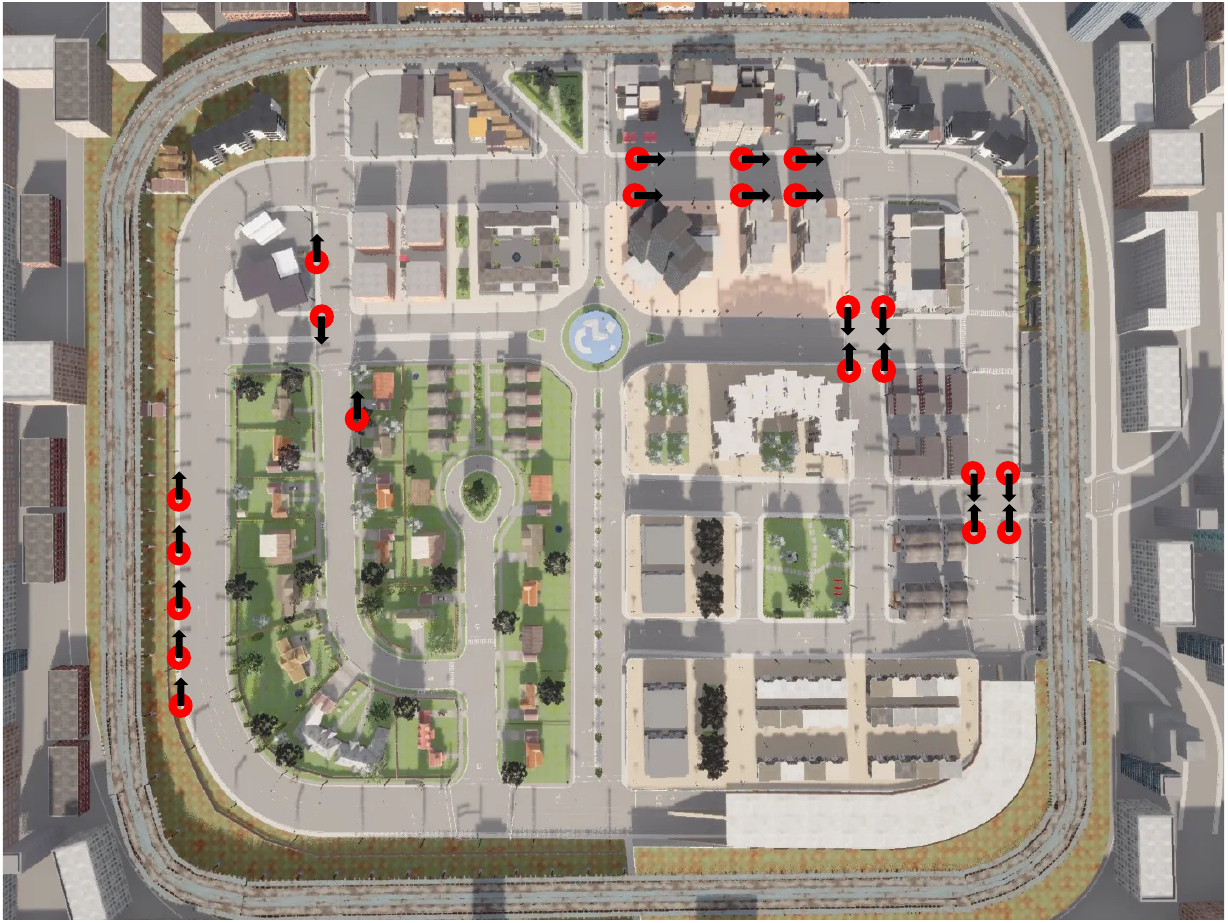}
    \end{subfigure}
    \caption{Overview of camera placement in Town 3, CARLA simulator. Red markers with black arrows indicate the location and direction of traffic cameras.}
\end{figure}

\end{document}